# Wavefront correction with a ferrofluid deformable mirror: experimental results and recent developments

Denis Brousseau*[a], Ermanno F. Borra[a], Simon Thibault[b], Anna M. Ritcey[c], Jocelyn Parent[a], Omar Seddiki[a], Jean-Philippe Déry[c], Luc Faucher[c], Julien Vassallo[a], Azadeh Naderian[a]

[a]Département de physique, génie physique et optique and COPL, Université Laval, Québec, Québec, CANADA G1V 0A6;
[b]Immervision, 2020 University St., Suite 2420, Montréal, Québec, CANADA H3A 2A5;
[c]Départment de chimie, CERSIM and COPL, Université Laval, Québec, Québec, CANADA G1V 0A6

## ABSTRACT

We present the research status of a deformable mirror made of a magnetic liquid whose surface is actuated by a triangular array of small current carrying coils. We demonstrate that the mirror can correct a 11 μm low order aberrated wavefront to a residual RMS wavefront error 0.05 μm. Recent developments show that these deformable mirrors can reach a frequency response of several hundred hertz. A new method for linearizing the response of these mirrors is also presented.

**Keywords:** Adaptive optics, liquid mirrors, deformable mirrors, aberration compensation

## 1. INTRODUCTION

The number of new adaptive optics applications and new very large telescope projects has soared during the last decade, demonstrating the need for low-cost, high stroke deformable mirrors with a large number of actuators.

Liquid Mirror Telescopes (LMTs) are good examples of the use of a liquid to make low-cost optical components. They have proven their usefulness in obtaining astronomical, space and atmospheric science data[1-5]. Besides cost, the fact that a liquid follows an equipotential surface to a very high precision is one of their major advantages.

Ragazzoni and Marchetti[6] discussed the use of liquids to build deformable mirrors in 1994. Their proposition consisted of passing currents through mercury and actuating the surface with a magnetic coil. However, the high density of mercury means that strong magnetic fields and thus high electrical currents are required. Liquid adaptive mirrors shaped by electro-capillary effects have also been discussed by Vuelban et al. in 2006[7].

The limitations associated with the use of mercury for liquid deformable mirrors can be solved through the use of ferrofluids[8]. Ferrofluids are suspensions of small (about 10 nm in diameter) ferromagnetic nanoparticles dispersed in a liquid carrier. In the presence of an external magnetic field, the ferromagnetic particles align with the field and the liquid becomes magnetized. Using proper magnetic field geometries, any desired shape can be produced at the surface of the ferrofluid. Following the pioneering work of Laird et al.[9], we have demonstrated that several ferrofluid surfaces matching optical aberration terms can be produced by means of current carrying networks configured as straight wires and/or loops[10, 11].

Ferrofluids have a reflectivity of about 4% and need a reflective coating for certain optical applications. Our group has a strong research history with coatings made of colloidal silver nanoparticles called MELLFs[8] (*Metal-Like Liquid Films*). Although MELLFs are not compatible with commercial ferrofluids, we have successfully developed our own ferrofluids having characteristics necessary for compatibility with MELLFs. Our team has also demonstrated that certain liquids can be coated using surface deposition techniques that could be adapted to ferrofluids[12].

This paper presents our first functional prototype of a ferrofluid deformable mirror and the current status of our research.

*denis.brousseau.1@ulaval.ca; phone 1 418 656-2131 ext. 4646

## 2. THE 37-ACTUATOR FERROFLUID DM

### 2.1 Description of the FDM

Our prototype FDM consists of 37 custom made coils (actuators) closely packed in a triangular array 35 mm in diameter (see Fig. 1). The choice of using a triangular geometry is based on previous simulations by Rioux[13] that demonstrated that such a configuration should provide better wavefront residuals than a square or circular arrangement. Each actuator is made of about 200 loops of resin-coated copper wire, 0.3 mm in diameter, and has an external diameter of 5 mm[14]. A small ferrite core, 1.5 mm in diameter, is placed at the center of each actuator to lower the current consumption to a maximum of 200 mA per actuator. The currents in the actuators are controlled by a LabView interface and three Adlink 6216V 16-channel analog output cards coupled with simple amplifier stages. An aluminum container filled with a one millimeter thick layer of EFH1 ferrofluid from Ferrotec (USA) Corporation is placed on top of the 37 actuators. EFH1 is an oil-based ferrofluid that has a relative permeability of 2.7 and a density of 1210 kg.m$^{-3}$. We did not use a MELLF coated ferrofluid as high reflectivity was not needed for our testing purposes.

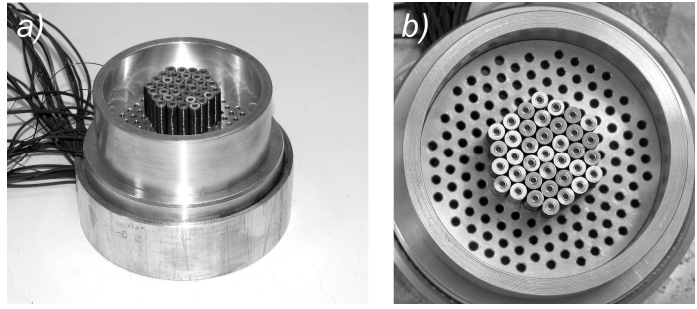

Fig. 1. (a) The 37-actuator ferrofluid deformable mirror with its aluminum support. b) Closer view of the mirror showing the 5 mm diameter actuators. The triangular array of actuators covers a diameter of 35 mm.

The cost of the FDM was estimated at about $100 per actuator. The estimated cost includes materials, electronics and shop time. About 70% of the total cost is due to the commercial analog output cards and could be decreased by using newer cards with a higher number of outputs.

### 2.2 Modeling of the FDM

The equation that describes the surface of a layer of ferrofluid lying in the *xy* plane in the presence of a magnetic field **B** is given by[15]

$$\zeta(x, y) = \frac{(\mu_r - 1)}{2\mu_r \mu_0 \rho g} \left[ |\mathbf{B} \cdot \mathbf{n}|^2 + \mu_r |\mathbf{B} \times \mathbf{n}|^2 \right] \quad (1)$$

where $\rho$ is the density of the ferrofluid, $\mu_r$ the relative magnetic permeability of the ferrofluid and **n** is a unit vector perpendicular to the surface of the ferrofluid. The maximum achievable surface amplitude is only limited by the so-called Rosensweig[16] instability and corresponds to a deformation of the order of a millimeter for the type of ferrofluid used in our experiments. For our FDM, the magnetic field **B** is the sum of all the individual magnetic fields produced by the actuators. The magnetic field of an actuator can be numerically computed under the assumption that it is composed of several current loops and that the ferrite cores only contribute as an amplifying factor[14]. However, Eq. 1 is non-linear with respect to the external magnetic field and depends on the vectorial addition of each magnetic field component produced by the actuators. This means that modal control of a FDM is impossible. We have successfully developed an algorithm that is able to compute the currents to supply to the actuators for a targeted mirror surface. This algorithm is similar to a regular zonal control but requires additional steps because of the need to individually consider each magnetic field component[14].

Using Eq. 1, our calculations shows that the response of a single actuator is well described by a Gaussian profile and that the FWHM of the deformation can be tuned by varying the distance between the ferrofluid surface and the summit of the actuators. We have computed the residuals for several Zernike polynomials for various distances between the ferrofluid and the actuators and found that a distance of 4 mm is optimal for this FDM. By analogy with conventional DMs, this corresponds to a coupling factor of about 25%. However, it must be noted that this coupling factor value is not directly comparable with conventional DMs specifications because of the previously discussed implications of Eq. 1.

**2.3 Results and example of static wavefront correction**

Fig. 2 illustrates the computed RMS wavefront residuals of the FDM for the first 32 Zernike polynomials following OSA convention[17]. Residuals in Fig. 2 are given as a fraction of the RMS amplitude of the targeted Zernike polynomial.

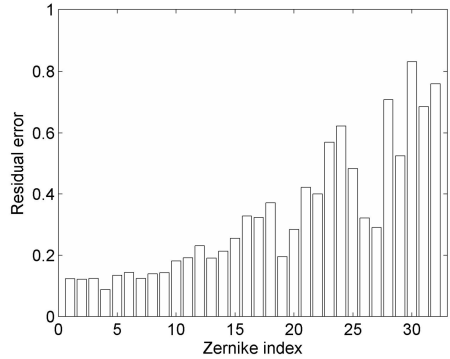

Fig. 2. Predicted residuals for the 37-actuator FDM. Residual errors are given as a fraction of the RMS amplitude of the targeted Zernike. OSA Zernike numbering scheme is used.

The residuals are below 0.3 up to the 16$^{th}$ Zernike term. Terms higher than 16 still provide about 50% wavefront correction, on average. Numerical simulations indicate that these residuals decrease by a factor of 2 when the number of actuators is increased to 127.

Computed currents for the first 20 Zernike polynomials were fed to the FDM and the resulting wavefronts were measured using an Imagine Optics HASO HR44 Shack-Hartmann wavefront sensor setup. Since no closed-loop control was implemented at that time, small dissimilarities between the actuators implied that small individual manual current adjustments were needed. These dissimilarities are caused by small differences in the exact number of windings of the actuators and inhomogeneities in the ferrite cores. It was found that the residual errors for low order modes were close to predicted values but 1.5 to 2 times larger than predicted for Zernike terms higher than 10. The higher spatial frequency of these terms renders the manual current adjustments necessary for troublesome actuators more difficult. Driving the FDM in a closed-loop configuration would certainly reduce residuals for these higher terms.

Because the response of a FDM is non-linear and shows a vectorial dependence on magnetic field strength, we suspected that correction of more complex wavefronts composed of several Zernike terms would give different residuals than the individual consideration of the residuals of each term present in the aberrated wavefront. We thus tested the FDM performance in two experiments in order to asset its ability to correct complex static wavefronts. One of the experiments consisted of purposely misaligning lenses in the wavefront measurement setup. The other experiment involved placing two small Petri dishes in the optical path of the measurement setup.

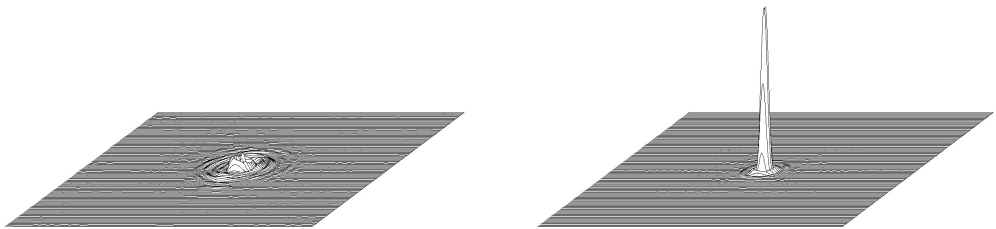

Fig. 3. PSFs before (left) and after (right) correction with the 37-actuator FDM. Wavelength used is 659.5 nm and Strehl ratio of corrected PSF is 0.84.

The misalignment experiment introduced spherical aberration and astigmatism in the system. The PV and RMS amplitudes were measured to be 2.42 and 0.58 $\mu$m, respectively. The required currents were computed and applied to the FDM, bringing the wavefront residual RMS error down to 0.05 $\mu$m. PSFs were computed from the measured coefficients and appear in Fig. 3. The Strehl ratio of the corrected wavefront is 0.84 at a wavelength of 659.5 nm.

The second experiment introduced much higher amplitude aberration terms. The measured PV and RMS wavefront amplitudes were found to be 11.43 and 2.58 $\mu$m, respectively. Currents were calculated and applied to the FDM, resulting in a residual wavefront RMS error of 0.15 $\mu$m. If we strictly consider the first 10 Zernike terms, the wavefront PV amplitude is still around 11 $\mu$m but the corrected wavefront RMS residual drops to 0.05 $\mu$m. The FDM is thus a great tool to compensate for high amplitude low aberration terms. Higher Zernike terms would have better residual wavefront errors with a FDM having a greater number of actuators.

## 3. RECENT DEVELOPMENTS

### 3.1 Linear control of FDMs

The 37-actuator FDM prototype has shown promising results but some drawbacks still remain. The amplitude of the deformations produced at the surface of the ferrofluid shows a non-linear dependence on the applied magnetic field (or current) produced by the actuators. Secondly, vectorial behaviour of the magnetic field prohibits the use of standard control algorithms to predict the surfaces produced by the FDM. Also, since deformations are proportional to the square of the applied magnetic field, only positive deformations can be produced on the FDM surface. Finally, although the ferrite cores employed inside the actuators reduce the current requirements, they render the predictability of the FDM performance more complex. Although it has not been experimentally observed, the ferrite cores might also suffer from hysteresis effects. Vibration susceptibility was demonstrated not to be a major issue[16] and recent developments (see section 3.2) that now favour the use of viscous ferrofluids corroborate this fact.

A new way to control FDMs that solves these drawbacks has recently been proposed[18]. By using a sufficiently large uniform magnetic field perpendicular to the surface of the ferrofluid, this new control method allows the amplitudes to be linearly dependent on the magnetic field of the actuators. This is because, in this configuration, the only term contributing to the surface amplitude is the one in which the small magnetic field of the actuator is multiplied by the large uniform magnetic field. Moreover, as the small magnetic field of an actuator is multiplied by a large uniform field, this reduces the current requirements for the actuators and permits the removal of the ferrite cores. This renders possible the fabrication of smaller actuators with a higher density of surface coverage. This linear behavior also allows for the calculation of surfaces produced by the FDM with regular modal or zonal control algorithms. Linearization of the FDM response also makes negative deformations possible ("push-pull" effect) and so doubles the available stroke of the mirror without increasing residual errors.

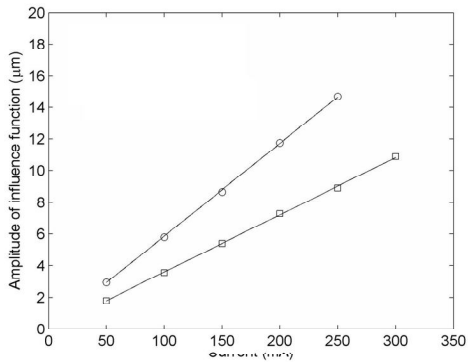

Fig. 4. Measured amplitude of the influence function of an actuator (without ferrite core) in the presence of a uniform external magnetic field from a Maxwell tricoil. The two linear curves correspond to values of external magnetic field of 20 and 30 gauss, respectively. Inversion of the currents produces negative amplitudes.

To test this method, we placed a single actuator (without a ferrite core) in the center of a Maxwell tricoil. A Maxwell tricoil is a derivative of the standard Helmholtz coil except it has a larger volume inside which the magnetic field is constant. Current is supplied to the Maxwell coil producing a constant magnetic field of about 25 gauss (about 10 times the value produced by an actuator). As can be seen in Fig. 4, our preliminary results confirm the linear response resulting from using this method. The response of the actuator is linear and much higher amplitudes are produced for lower currents than those required in the absence of the Maxwell tricoil. As can be seen in Fig. 4, the response of the actuator can be amplified by increasing the external constant magnetic field. The deformations produced are sufficient to correct aberrations present in the abnormal human eye and allow the mirror to be used as an optical test element.

### 3.2 Dynamic response of FDMs

Early experiments indicated that FDMs would be limited to dynamic uses at frequencies lower than 20 Hz. When driven at frequencies higher than this cut-off frequency, there is a rapid loss in amplitude of the liquid's response is observed and a phase lag of over 90 degrees appears between the driving electric signal and the output ferrofluid deformation. These observations indicate that the bandwidth of a FDM is comparable to that of a low-pass RLC filter.

We found that the amplitude loss at high frequencies can be countered by overdriving the actuators. Usual driving of the mirror consists of feeding each actuator a step function whose amplitude is proportional to the targeted deformation. The overdriving technique resides in adding a very short and high amplitude current pulse at the beginning of each step function. Because the characteristic rise time of the ferrofluid is independent of the deformation's height, the overdriving pulses give a higher initial velocity to the liquid. The desired surface amplitude is obtained faster and the remaining part of the step function stabilizes the liquid.

It thus seemed that the main obstacle with high frequencies was the phase lag. The dynamic response of a ferrofluid is much like the response of a driven damped harmonic oscillator. At low frequencies, there is almost no phase lag between the input signal and the output response. A phase lag of 90 degrees is observed at the system's natural frequency and, at higher frequencies, the phase lag approaches 180 degrees.

In more recent experiments, however, we found that the phase lag can be countered by increasing the viscosity of the liquid. The frequency for a critical phase-lag of 90-degree was improved from 20 Hz to 450 Hz by increasing the viscosity of the ferrofluid from 6 cP to 494 cP. Since a single period of a square wave corresponds to two corrections, this actually implies a 900 Hz bandwidth. This mirror bandwidth value must not be confused with the actual adaptive optics system bandwidth, which is also slowed down by the finite speeds of the wavefront sensor and the wavefront reconstructor. However, as can be expected that by increasing the viscosity, the time required for stabilization would be longer. The liquid now acts as an over-damped harmonic oscillator. Here again, the solution to this problem is to use overdriving pulses that give an initial velocity to the liquid.

To summarize, it is possible to counter the problematic phase lag at high frequencies by increasing the viscosity of the liquid. However, increasing viscosity and frequency decreases the amplitude reached by the liquid. This effect can be countered by using overdriving pulses to drive the mirror. With these two modifications to the mirror, using FDMs in closed-loop at a running frequency of 500 Hz appears to be realistically attainable.

## 4. CONCLUSION

We have demonstrated that a prototype 37-actuator ferrofluidic deformable mirror can correct an 11 $\mu$m static wavefront error to a RMS wavefront residual error of 0.05 $\mu$m. A Strehl ratio of 0.84 has been achieved using this FDM. Although we were able to model the surfaces produced by the mirror, implementation of standard DM control algorithms remained impossible. A new way to control these FDMs was introduced and uses a constant magnetic field surrounding the mirror that allows the response of the actuators to become linear. This makes possible the use of regular zonal or modal control of the mirror and these methods are now being investigated by our group. This also simplifies our goal of demonstrating closed-loop operation of FDMs.

Our experiments have shown that using a ferrofluid having a sufficiently high viscosity enables us to reach a bandwidth of 900 Hz. This bandwidth value opens the door to many more applications than previously thought. Tests on the chemical deposition of thin membranes that could be coated with a reflective layer are also underway and could also help to improve the bandwidth of FDMs.

The prototype FDM presented is made of standard components and was build at a total cost of about $100 per actuator, including parts and labor in our mechanical shop. With a better engineered mirror, costs could be further reduced. The low-cost aspect of FDMs combined with the large deformations achievable with ferrofluids, suggests that low-cost deformable mirrors having a very large number of actuators are possible. Ongoing research shows that the diameters of

the actuators could be reduced to about 1 millimeter, thus allowing for a greater packing density. FDMs show promise for an interesting application in which a low-cost and low optical quality primary mirror having a simple and inexpensive supporting mount is used with active correction by an FDM, having a very large number of actuators that corrects the poor optical quality of the primary mirror. This application is made possible by the low costs and large amplitudes of FDM mirrors. The combination of an inexpensive primary and support system with an FDM would thus allow us to build inexpensive optical telescopes.

Both the FDM prototype presented here and the recent developments described in this paper are very encouraging and predict a great future for this new kind of deformable mirror.

## ACKNOWLEDGEMENTS

This research was supported by the Natural Sciences and Engineering Research Council of Canada and NanoQuébec.